\documentclass[useAMS,usenatbib]{mn2e}
\usepackage{graphicx}
\usepackage{txfonts}
\begin{document}

\title[Single pulse analysis of PSR B1133+16 at 8.35 GHz and carousel circulation time]
{Single pulse analysis of PSR B1133+16 at 8.35 GHz and carousel circulation time}
\author[Sneha Honnappa et al.]
{Sneha Honnappa,$^1$\thanks{E-mail: sneha@astro.ia.uz.zgora.pl}
Wojciech Lewandowski,$^1$
Jaroslaw Kijak,$^1$
Avinash A. Deshpande,$^2$
\newauthor
Janusz Gil,$^{1}$
Olaf Maron$^1$
and Axel Jessner$^3$ \\
%EndName
$^1$ Kepler Institute of Astronomy, University of Zielona G\'ora , Lubuska
2, 65-265 Zielona G\'ora, Poland\\
$^2$ Raman Research Institute, Bangalore 560 080, India\\
$^{3}$ Max Planck Institute f\"ur Radioastronomie, Auf dem H\"ugel 69, D-53121 Bonn, Germany}
\date{Accepted . Received ; in original form }
\maketitle

\begin{abstract}
A successful attempt was made to analyse about 6000 single pulses of PSR B1133+16 obtained with the
100-meter Effelsberg radio--telescope. The high resolution (60 micro-seconds) data were taken at a
frequency of 8.35 GHz with a bandwidth of 1.1 GHz. In order to examine the pulse-to-pulse
intensity modulations, we performed both the longitude-- and the harmonic--resolved
fluctuation spectral analysis. We identified the low frequency feature associated with
an amplitude modulation at
$f_4\simeq 0.033 P_1^{-1}$, which can be interpreted as the circulation time $P_4\simeq 30 P_1$ of
the underlying subbeam carousel model. 
Despite an erratic nature of this pulsar, we also
found an evidence of periodic pseudo-nulls with $P_4 = 28.44 P_1$. This is exactly the value at
which Herfindal \& Rankin found periodic pseudo-nulls in their 327~MHz data. We thus believe that
this is the actual carousel circulation time in PSR B1133+16, particularly during orderly circulation.
\end{abstract}

\label{firstpage}


\section{Introduction}

Generally pulsars are weak radio sources, and even with the sensitivity of current radio telescopes
single pulses are observable only from the strongest objects. Due to the steep spectrum of pulsar
radio--emission, single pulse observations were readily made at low frequencies, where the
sensitivity requirements for antenna and receiver systems were not an issue. PSR B1133+16 is one of
the brightest pulsars in the northern hemisphere \citep[32~mJy at 1.4~GHz, spectral index of $-1.5$][]{maro00}. 
Individual pulses of PSR~B1133+16 were recorded during numerous observing
campaigns, and analyses of single pulse behaviour were performed at the observing frequencies
from 110~MHz to 1.4~GHz \citep{Backer,Taylor,Nowa,Welte,Weltevrede,Herfindal}. With the recent
development of high sensitivity receivers, and given the fact that PSR~B1133+16 is a relatively
strong source, measurements of its single pulses with high time resolution were possible at
frequencies as high as 8.35 GHz.

The most common feature revealed by single-pulse sequences is related
to the phenomenon of drifting subpulses \citep{Drake,Sutton,Weltevrede}. The parameters of the
drift were first described by \citet{Backer}, later modified by \citet{Gil03} and \citet{Gupta04}.
The first of the four main periodicities involved is the basic pulsar (spin) period  $P_1$,
i.e. the interval between our successive sampling of the pulsar beam.
The overall pulsar beam is organized into a number of subbeams, only some of which cross our sightline
during any given pulse, and thus directly correspond to the number and the locations of subpulses 
we observe.
The second periodicity, $P_2$, is the angular distance
between adjacent subpulses, or the corresponding subbeams,
as viewed along rotational longitude. The magnetic azimuthal separation, say $\eta$,
between adjacent subbeams implied by $P_2$ can be estimated \citep[e.g., see Eq. 3 of][]{Deshpande}.
For $N$ subbeams, when evenly placed within the emission cone, $\eta$ would be $= 360^{\circ}/N$. 
The next period $P_3$ (usually measured in
units of $P_1$, and as per its initial definition) is the distance between adjacent drift bands, 
or in other words, the interval after which the subsequent subbeam reaches the same phase as 
the preceding one. 
In case of fast drifts, however, one has to take into account possible aliasing effect, which can cause a
large degree of ambiguity. Fluctuation spectral analysis can help to resolve the aliasing issue, and
is therefore commonly used. In such Fourier
analysis, the periodic phase modulation associated with the subpulse drift manifests as
a spectral feature at the corresponding fluctuation frequency, $f_3\equiv 1/P_3$. This feature
is usually seen at relatively high fluctuation frequencies, and hence, often
called the high frequency feature (HFF). Finally, we can introduce the total
circulation time of the pattern, or the interval after which a given subbeam reaches the same
phase, $P_4$ \citep[$\hat{P}_3$ in ][ RS75 henceforth, who introduced this feature
theoretically]{rs75}. The circulation time becomes measurable if the subbeams vary in intensity, 
as they most often do, and such periodic amplitude modulation would be apparent as an accompanying 
spectral feature at frequency $f_4 \equiv 1/P_4$.
This spectral feature, when present, is invariably located in the lower band of the 
fluctuation spectrum, and is referred to as the low frequency feature (LFF). 
According to the carousel model,
first introduced by RS75 and developed by \citet{Desh,Deshpande} and \citet{Gil00,Gil03}, the two
periodicities $P_3$ and $P_4$ are connected to each other via a direct relationship  $P_4 = N P_3$.

PSR~B1133+16 apparently shows the phenomenon of drifting subpulses, although it was shown that the 
organized character of the subplulses appears only for some finite timespans, outside of which
the behaviour of the individual pulses is highly chaotic. Recently \citet{Weltevrede}
detected the high-frequency (phase-modulation) spectral feature corresponding to period of $P_3 = (3
\pm 2) P_1$. Moreover, they also detected a low-frequency (amplitude-modulation) feature
corresponding to period $P_4 = (30\pm 8) P_1$. The very large uncertainty in the former estimate follows
from a fact that the authors were not able to resolve the issue of aliasing in their analysis.
\citet[HR07 henceforth]{Herfindal} attempted to resolve the aliasing problem and demonstrated that the
actual $P_3 = (1.237\pm 0.011) P_1$. Their spectral and profile-folding analyses confirmed the
existence of the low-frequency feature (LFF) in the spectrum, corresponding to a period of $P_4 =
28.44 P_1$. The unprecedented accuracy of the latter estimate is a result of sophisticated
analysis, including the cartography technique developed by \citet{Desh,Deshpande}, as well
as detection of periodic pseudo-nulls of very short duration. It is worth noting that the value of
N=28.44/1.237 is exactly 23 in their analysis, as it supposed to be for a parameter describing the
number of subbeams contributing to the observed drifting subpulse phenomenon. Thus, the value of
$P_4=28.44 P_1$ represents the carousel circulation time in PSR B1133+16 (HR07).

In this paper we attempt to find the pulse modulation features using spectral analysis of our high
frequency (8.35~GHz), high time resolution (60 $\mu$s) single pulse data of PSR~B1133+16.

\section{Observations and Data Analysis}

%\subsection{Observations}

Our observations of single pulses of PSR~B1133+16 were made with the 100-meter radio telescope
of Max-Planck Institute for Radioastronomy in 2004. We used a cooled secondary-focus receiver
equipped with HEMT amplifiers, with an observing frequency of 8.35 GHz and 1.1 GHz bandwidth.
The pulsar signal was recorded by Effelsberg Pulsar Observations System \citep[EPOS,][]{Jessner}, with a
time resolution of 60 micro-seconds. To achieve the best pulse coverage, we used a gating system, in
which 1024 bins were used to cover only about 5\% of the pulse period around the main pulse. Data were
gathered during two separate 60-minute sessions (c.a. 3000 pulses each).

Figure~1 shows a sample of the data we gathered; a set of 256 consecutive individual
pulses, with its average pulse profile at the top. One can see that only some of the individual
pulses were clearly detected. For a majority of the pulses, the radio emission was very weak or
even missing. Naturally, the typical phenomena related to subpulse drifting, like the appearance of
drift bands or high-frequency features in the modulation spectra, may not be readily apparent.
However, contrary to our low expectations, we managed to detect all periodicities
appearing in the pulse intensity modulations visible in better quality low-frequency data. The level of
confidence of these detections varied between different segments of the data set. For example, a
low-frequency amplitude-modulation feature was very prominent in the analysis of one set of pulses,
while in some sets the feature can be recognized in the spectrum, but not at a level sufficient 
to claim a clear detection. However, the particular set (segment B; pulses 512-768) shown in Figure~1
is one of our best-quality intervals, where the periodic  
intensity modulation, with $P_4$ close to 28~$P_1$, is clearly visible.

\begin{figure}
\includegraphics[width=11.0cm,height=9.0cm,angle=-90]{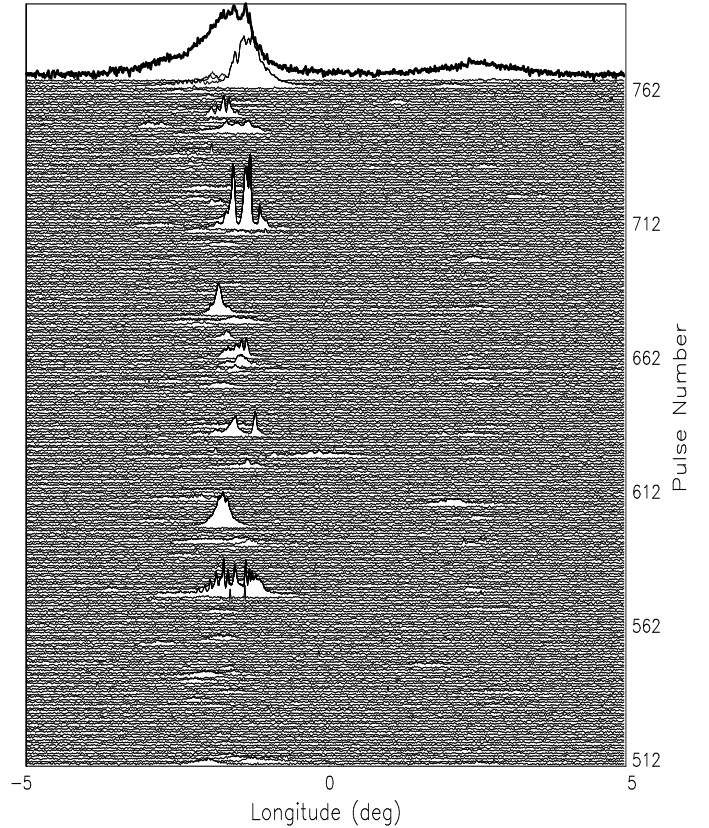}
\caption{A sequence of pulses from 512th to 768th pulse from observation B, showing clearly a periodic 
intensity modulation (conf. Tables~1~and~2). The thick line at the top represents the integrated profile 
for this sequence. }
\end{figure}

\subsection{Longitude--Resolved Fluctuation Spectra}

In order to analyse the gathered data, we used the method of Longitude-Resolved Fluctuation
Spectra \citep[LRFS hereafter,][]{Back}, which detects the regularities in the intensity variations
of individual pulses. To implement the method, we divided the pulse series into blocks of 256
pulses, and performed the one-dimensional fast Fourier transform (FFT) for each longitude bin, and
computed the power spectrum. The spectra were then normalized for each individual longitude bin,
resulting in 1024 independent power spectra. These were then summed to form an integrated power
spectrum separately for each of the 256-pulse stacks.

A feature with periodicity around 2.85 $P_{1}$ was identified in the fluctuation
spectra at {\it all} longitudes, as well as in the integrated spectrum.
This feature was present in all our
pulse sequences, and it's origin was traced down to the AC power line. We confirmed that this
feature was not associated with the pulsar signal by obtaining LRFS for the off-pulse region. 
The feature was extremely strong, potentially masking any real features present in the contaminated part of
the spectrum. Therefore, we decided to
remove it from the fluctuation spectra. To achieve this, we reconstructed a continuous time series from the
available gated data and calculated its FFT. In this fluctuation spectrum, we replaced the values
at the locations corresponding to the fundamental frequency of the contaminant, 
and its harmonics (separated by the pulsar frequency as a consequence of the gating), with zeros. 
An inverse Fourier transform of thus {\it edited} spectrum
yielded a modified time series, free of this interference, which after the re-application of
gating allowed us to repeat the LRFS analysis.

\subsection{Harmonic resolved fluctuation spectra}

The LRFS gives us a wealth of information about the fluctuations
present in the sequence, but also has a serious drawback of aliasing due to poor sampling of 
fluctuations at a given longitude only once every pulse period. In order to deal with the issue of
aliasing, we need to overcome the basic limitation of this sampling. Noting that the fluctuations
are sampled more than once in the finite duration of the pulse, one can combine the fluctuation spectra of
the different longitude bins with appropriate phases corresponding to the longitude separation to
obtain the Harmonic-Resolved Fluctuation Spectra (HRFS hereafter). This approach was used and
developed by \citet{Desh,Deshpande}. In practice, this is achieved by Fourier transforming the
entire continuous sequence, which can be reconstructed (from the gated version) by using the
available on-pulse samples, and filling the rest of the longitude regions with zeros. In this
process, the pulse sequence is effectively multiplied by a periodic window function which leads to
convolution of the spectrum associated with the un-gated pulse sequence with the Fourier
transform of the periodic window (gating) function used.

\section{Results}
\label{results}
 We analysed several 256-pulse (c.a. 300 second) data blocks from both of our 60-minute
observations (we call them $A$ and $B$ for the purposes of referencing here). For some
of these blocks, we have identified a prominent low frequency feature (LFF) in the integrated
spectra obtained with LFRS, as well as HRFS, analysis.

Figure~2 shows the result of our LRFS analysis for an interference-cleaned 256-pulse
stack, and Table~1 lists the values of the measured frequency $f_4$ of the LFF and the corresponding
periodicity $P_4=1/f_4$ for those data blocks in
which these features were detected using the LRFS analysis. For the remaining data blocks the LFF
was either not present at all (no identifiable peaks close to this frequency in the power spectrum),
or its signal-to-noise ratio was too small to claim its presence beyond doubt
(usually due to other features that appeared in the power spectrum). The $f_4$
values corresponding to the LFF 
vary between the blocks, but stay consistent within the estimated uncertainties.

For the observation $A$, it was possible to identify the LFF in the average spectrum,
obtained as a sum of the spectra from all the data blocks. The value of the LFF frequency from the
average spectrum (see Table 1) was close to the $28.44 P_1$ obtained by HR07
as the putative circulation time of their carousel solution. The same was not possible 
for observation $B$, as the
LFF frequency varies significantly from block to block, resulting in a smearing of the feature in
the averaged spectrum. However, it is worth noting that in the best block B (512-768, see Fig.~1) with the
lowest estimated errors, the $P_4=(28.4\pm 1.2) P_1$ was also very close to the
HR07 value. Table~1 also contains the weighted average of the LFF period,
obtained from all individual block measurements.

The result of our HRFS analysis of the same 256-pulse sequence (as used for the LFRS in Fig.~2) 
is shown in Fig.~3.
The HRFS for these pulse sequences also show the LFF,
usually with another feature (at the frequency equal to 1.0 minus the LFF frequency; conf. Figure~3),
which would be expected to be present as the LFF's symmetric counterpart (as in a pair of sidebands
about each of the pulsar harmonics) in case of amplitude modulation. The results of the measurement
of the LFF parameters (see Table~2) agree with those from the LRFS analysis. 
Similar to the case with LRFS,
we were able to find the LFF in the average HRFS only for the observation $A$, and the estimated 
weighted-average frequency (and the corresponding period) associated with the LFF are also presented in Table~2.

\begin{table}
\caption{Parameters of the observed low-frequency feature (LFF) obtained from the LRFS analysis. Table includes also
the value obtained from the average spectrum of the whole observation A, as well as the weighted average of the LFF 
frequency, and the value of $P_4$ based on it (in parentheses).}
\small{
\centering
\begin{tabular}{c c c c c}
\hline\hline
Observation       & Pulse      &  \multicolumn{2}{c}{LFF}          \\
                  & sequence                   &  Period $[P_{1}]$ & Freq. $[P_1^{-1}]$           \\
\hline
A              & 1-256              & 28.4$\pm$3.3  &  $0.035\pm0.004$     \\
A              & 512-768            & 28.4$\pm$3.5  &  $0.035\pm0.004 $     \\
A              & 2109-2365          & 32.0$\pm$2.2  &  $0.031\pm0.002$     \\
B              & 1-256              & 36.5$\pm$4.5  &  $0.027\pm0.003$      \\
B              & 512-768            & 28.4$\pm$1.2  &  $0.035\pm0.002$     \\
B              & 2304-2560          & 34.6$\pm$9.5  &  $0.029\pm0.007$       \\
\hline
\multicolumn{2}{c}{Freq. weighted average} & (30.9$\pm$1.1)   & $0.0323\pm0.0012$      \\
\hline
\multicolumn{2}{c}{Average spectrum of A$^{\star}$} & 28.3$\pm$7.2  &  $0.035\pm0.009$      \\
\hline
\multicolumn{4}{l}{$^{\star}$ - see text in Section~\ref{results} for details.}\\
\end{tabular}
}
\end{table}

\begin{figure}
\resizebox{\hsize}{!}{\includegraphics{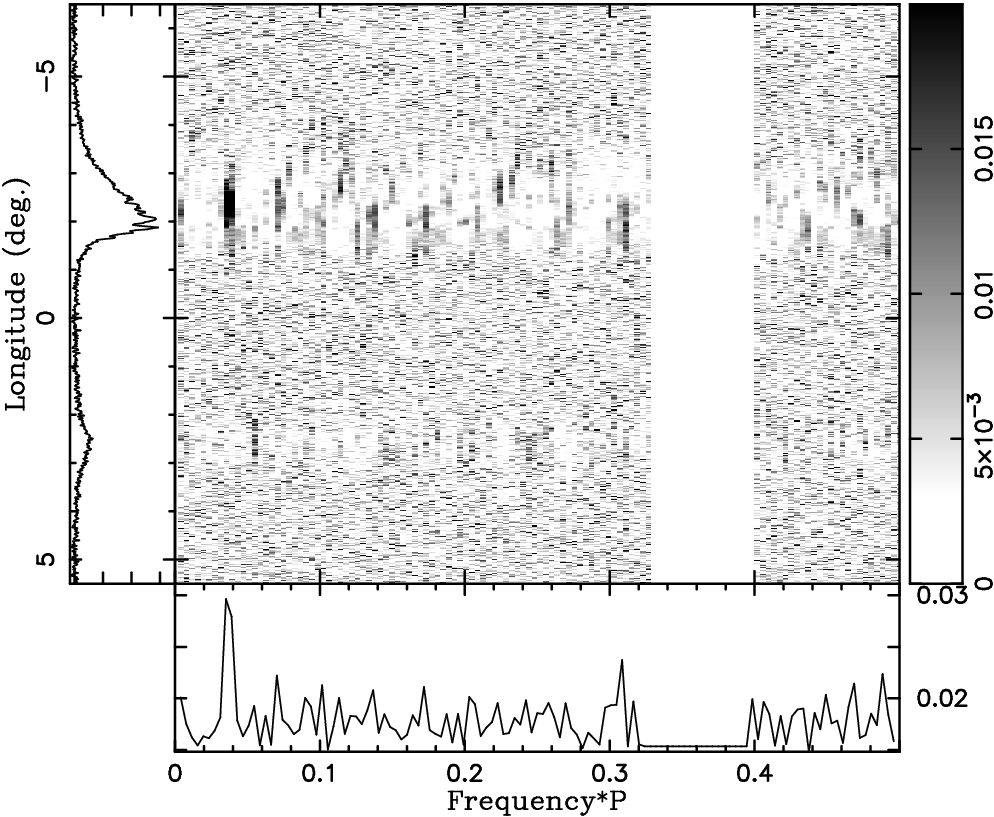}}
\caption{Longitude-resolved fluctuation spectrum for PSR B1133+16 computed for 8.35 GHz 
observations (using the sequence from 1st to 256th pulse from observation A). 
The main panel presents, in grey-scale, the full fluctuation spectrum for all longitude bins 
corresponding to the integrated pulse profile, which is shown in the left hand panel. 
The bottom panel presents the integrated fluctuation spectrum; note the prominent feature 
close to 0.03$P_1^{-1}$.
This feature is clearly visible also in the LRFS corresponding to the leading bright pulse component.
The white band visible in the LRFS greyplot is a result of removal of a whole range of frequencies that were
distorted by AC power line interference.}
\end{figure}

\begin{table}
\caption{Parameters of the observed low-frequency feature (LFF) obtained from the HRFS analysis. Table includes also
the value obtained from the average spectrum of the whole observation A, as well as the weighted average of the LFF 
frequency, and the value of $P_4$ based on it (in parentheses).}
\small{
\centering
\begin{tabular}{c c c c c}
\hline\hline
Observation            & Pulse      & \multicolumn{2}{c}{LFF}        \\
                & sequence   & Period $[P_{1}]$ & Freq. $[P_1^{-1}]$  \\
\hline
A            & 1-256              & 28.4$\pm$3.1 & 0.035$\pm$0.004\\
A            & 512-768            & 32.0$\pm$4.6 & 0.031$\pm$0.004\\
A            & 2109-2365          & 32.0$\pm$2.3 & 0.031$\pm$0.002\\
B            & 1-256              & 36.5$\pm$6.3 & 0.027$\pm$0.005\\
B            & 512-768            & 28.4$\pm$3.0 & 0.035$\pm$0.004\\
B            & 2304-2560          & 32.0$\pm$8.8 & 0.031$\pm$0.010\\
\hline
\multicolumn{2}{c}{Freq. weighted average} & (31.5$\pm$1.4) & 0.0317$\pm$0.0014\\
\hline
\multicolumn{2}{c}{Average spectrum of A$^{\star}$}& 25.6$\pm$3.3 & 0.039$\pm$0.005\\
\hline
\multicolumn{4}{l}{$^{\star}$ - see text in Section~\ref{results} for details.}\\
\end{tabular}
}
\end{table}

\begin{figure}
\resizebox{\hsize}{!}{\includegraphics{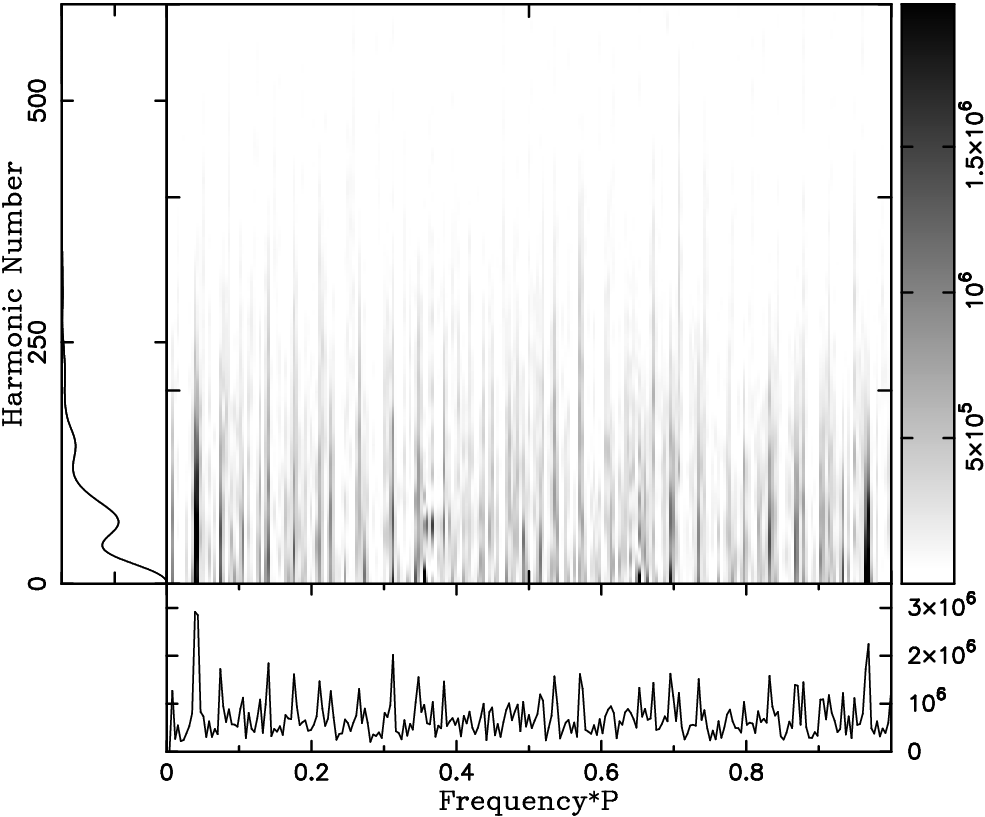}}
\caption{Harmonic-resolved fluctuation spectrum for PSR B1133+16 computed for the 8.35-GHz observations. 
The grey-scale plot in the central panel shows the full spectrum, 
and the bottom panel - the integrated spectrum. In both of them,
two features clearly stand out - one close to $0.03P_1^{-1}$, which is the low frequency feature
(LFF), and its symmetric counterpart at $\sim 0.97P_1^{-1}$.
The contributions, shown in the main panel, corresponding to any given frequency appear {\it stretched}
along the harmonic-number axis. This is an anticipated consequence of periodic windowing, including
the natural pulse-shape mask and any deliberate windowing, such as gating.
Left panel shows the relative 
strengths of pulsar frequency harmonics (equivalent to the Fourier transform of the average
pulse profile). 
The presented spectrum was
obtained for the same sequence of 256 pulses, as used for the LFRS in Fig.~2 (i.e. the first 256 pulses 
from observation A).
One has to remember that, for the purpose of HRFS analysis,
we had to reconstruct a continuous time series from the available gated observations.}
\end{figure}

\section{Discussion and Conclusion}

The average fluctuation frequency of the LFF , that we identified (in selected segments of our data), 
corresponds to a period of $(31.5 \pm 1.4) P_1$ (see Table~2).  We have chosen the value calculated
by the means of HRFS analysis as the best representative for a number of reasons (although one has to note that 
values from LRFS and HRFS agree well within the error estimates). First of them is that HRFS  adds the 
contribution associated with the modulations potentially coherently, using the implicit connection 
between modulation seen at different longitudes, as against the incoherent addition of modulations 
contribution across longitudes in LRFS. The former therefore is expected to provide a higher 
signal-to-noise ratio for coherent modulations, than the latter case. Also, while in case of HRFS the spectral 
power is divided between the feature and its alias, for amplitude modulated features the same goes for LRFS
analysis, we just do not look at the negative part of the spectrum. And finally, having two separated, but 
related features (alias frequency is 1 minus LFF frequency) allows for more precise measurement of the LFF 
position in the spectrum.

The value we obtained is consistent with other 
single-pulse studies of PSR B1133+16 (see Table~3). More importantly, a few segments of the 
best quality data show $P_4$ close to $28.44P_1$, the value given by HR07. They estimated this 
refined value of $P_4$ using profile folding over the circulation period, which is desirably 
sensitive to periodic pseudo-nulls.

Our data do not show a direct evidence of the HFF corresponding to subpulse drift
because of the erratic nature of the individual pulses (see Fig. 1); this is quite normal behaviour for 
this pulsar, which is - at our observing frequency of 8.35~GHz - amplified by the fact, that most of 
the incoming pulses remain below our detection threshold. Since $P_3$ (aliased or not)
measures the distance  between subpulse drift bands, it is clear that in our data, where most of
the pulses are missing, this periodicity cannot be easily detected. In addition one has to remember, 
that even at lower observing frequencies, the drifting subpulse phenomenon is not always present.
However, one should not ignore a putative drift feature at frequency about 0.81 $P_1^{-1}$, that 
we found in selected segments of our data (hidden in a forest of features in the average HRFS spectrum),
 which corresponds to the periodicity of about $P_3=1.237 P_1$. We note its similarity to
Fig.~A1 in HR07, where this drift feature at frequency about 0.81 $P_1^{-1}$ is clearly present.
It is also worth noting that $P_4/P_3=28.44/(1.23\pm 0.01)$ is exactly 23, which represents the number of
circulating beamlets in the carousel model of HR07.

There is no fundamental reason why the LFF corresponding to amplitude modulation
{\it also} should be weak or even missing in our data. Even when the sub-pulse phase behavior is less
ordered, and the primary phase fluctuation feature (HFF) may not be apparent in the spectrum, the
low-frequency feature indicative of circulation period may stand out clearly representing the
overall intensity variation across magnetic azimuth. The first of such examples of LFF was
discussed by \citet{Asgekar01,Asgekar05} based on their low-frequency (35 MHz) data of
B0943+10 and B0834+06. Also, a very good example is the so-called Q-mode in PSR~B0943+10 analysed
by \citet{rs06} in their 327-MHz Arecibo data. Unlike in the regular B-mode of this
pulsar with driftbands clearly visible and the HFF appearing in the fluctuation
spectrum, single pulses in the Q-mode are completely chaotic, and no driftbands appear in pulse
sequences \citep[see Fig.~4 in][]{rs06}. Still, in  this mode, the LFF
corresponding to carousel circulation time is clearly visible in the fluctuation spectrum
\citep[see Fig.~6 in][]{rs06}. Yet another example of pulsars with erratic subpulses, but
nevertheless showing the LFF that can be interpreted as the carousel circulation
time, is PSR B0656+14 (see footnote \ref{foot1}).

It is interesting and important that we are able to identify the LFF in our erratic data 
using both, LRFS as well as HRFS analysis. The LRFS 
does not, however, resolve the issue of whether the apparent spectral feature is indeed the true one or 
is an alias of another feature outside of the unaliased frequency range. The HRFS, on the other hand,
can help resolve such ambiguity to a large extent, and therefore we believe the estimate of the
frequency from the HRFS to be more reliable. Moreover, the HRFS 
analysis provides additional clues about the nature of modulation.

Previous studies reported an LFF at $f_4\simeq 0.033 P_1^{-1}$ (see
references in Table~3). In our HRFS spectra (see Fig.~3), we also clearly see this feature as the
most prominent one. There is also another significant feature at frequency $0.967 P_1^{-1}$, 
which is certainly the counterpart of the LFF ($0.033 P_1^{-1}$) \citep[see also Fig.~A1 in][]{Herfindal}. 
We thus believe that the true frequency $f_4$ is indeed about 0.033~$P_1^{-1}$, and 
the roughly similar intensities (within uncertainties) of the two spectral features are to be 
interpreted as suggesting significant amplitude modulation rather than
phase modulation (which would be the characteristic of drifting subpulses).
%In other words, subpulse drifting does not appear significant, which is consistent with
%its not being also apparent in the pulse sequence (see Fig.~1), possibly due to a combination of reasons
%including the poor signal-to-noise ratio during most of the sequence and the high intensities lasting
%only for a few pulses at a time. 
In any case, the prominence of the LFF in comparison with its harmonics indicates a 
rather systematic (non-random) and possibly smooth variation of intensity
across the pattern of subbeams.

\begin{table}
\caption{Results of observed Low Frequency Feature in PSR B1133+16 found in the literature and 
from our analysis}
\small{
\centering
\begin{tabular}{c c c}
\hline\hline
References               & Observation           &     Period from LFF      \\
                             & frequency         &   (in units of $P_{1}$)      \\
\hline
Slee \& Mulhall (1970)        &  80 MHz         &   37$\pm15$ \\      
                             &                   &             \\
Karuppusamy, Stappers        &  110 - 180 MHz        & 30$\pm$5        \\
  \&  Serylak (2011)            &                       & $34^{+5}_{-3}$ \\
                             &                       &  \\
Taylor \& Hugenin (1971)      &  147 MHz              &  27.8 \\
                             &                       &             \\
Weltevrede et al. (2006)     &  326 MHz              &  $30\pm8$\\
                             &                       &   \\\
Herfindal, Rankin (2007)     &  327 MHz              & 32$\pm$2  \\
                             &                       &              \\
Nowakowski (1996)            &  430 MHz              & 32 \\
                             &                       &     \\
Backer (1973)                &  606 MHz              & 37 \\
                             &                         &    \\    
Weltevrede et al. (2007)     &  1.4 GHz              & 33$\pm$5   \\
                             &                       &                          \\
Our Analysis (HRFS)          &  8.35 GHz             & 32$\pm$2 \\
\hline
\end{tabular}
}
\end{table}

\citet[GMZ06 henceforth]{Gil} claimed that, in the case of PSR~B1133+16, the LFF reported in the literature
with the period of $(33\pm 3) P_1$ corresponds to the carousel circulation time $P_4$.  
This claim was based on their Partially Screened Gap model
\citep{gmg03}, which predicts a precise correlation between $P_4$ and X-ray thermal luminosity
$L_x$ from the hot polar cap heated by back flow bombardments - see Eq.3 in GMZ06, and for more
details see \citet{Gil08}. Based on the available X-ray {\it Chandra} observations of B1133+16
\citep{Karg06}, GMZ06 found the predicted value of $P_4=\left(27^{+5}_{-2}\right) P_1$, well within
the expected range (see Table~3)\footnote{\label{foot1} Based on the same ideas, \citet{Gil07}
interpreted successfully the period  $(20\pm 1) P_1$ \citep[found in the fluctuation spectrum of PSR
B0656+14 by][]{Welte} as the carousel circulation time $P_4$ in this pulsar. This was an
especially attractive interpretation as the thermal X-ray emission in this case obviously
originated at the hot polar cap heated by backflow bombardment \citep{Deluca}. But what is even more important for
us, the radio single pulses of PSR B0656+14 looked almost exactly as erratic as those of PSR
B1133+16 at 8.35~GHz presented and analysed in this paper.}.

This interpretation of $P_4$ was later confirmed independently by HR07, although the authors
admitted that they strongly doubted it at first. However, they surprisingly found this periodicity
with a value of $P_4 = 28.44 P_1$  in data with very short duration nulls, which they called
pseudo-nulls. Such nulls can occur when the line of sight misses the emission between adjacent
subbeams (empty sightline traverses through the beamlet pattern that occur occasionally but
periodically). For the model which GMZ06 proposed, the value of $P_4$ should not depend on the
observing frequency. Our measurements for all available segments of data seem to support that
claim, as the LFF-based periods we measured are very much consistent (within their errors) with 
those found from pseudo-nulls at lower observing frequencies. This means that the carousel
circulation time stays constant over the observing frequencies ranging from 110~MHz to 8.35~GHz,
almost two orders of magnitude!

\begin{figure}
\resizebox{\hsize}{!}{\includegraphics{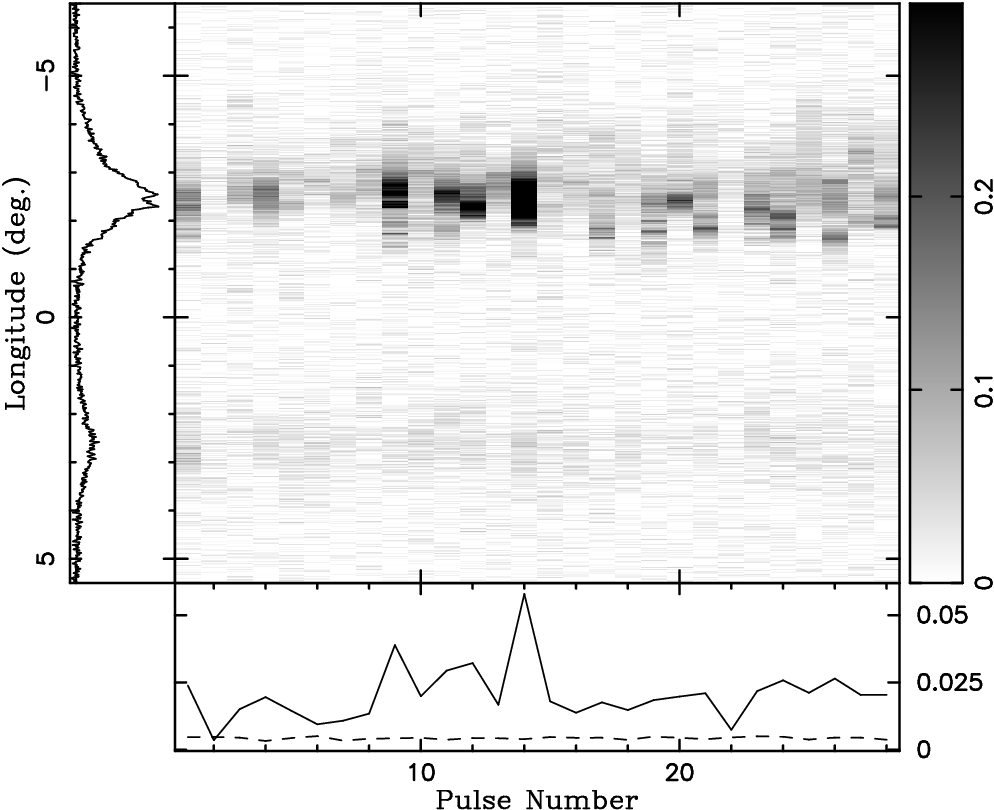}}
\caption{An example of the longitude-resolved modulation pattern (in grey-scale), 
obtained from 512 pulses folded at the putative carousel rotation period of $28.44 P_1$,
is shown in the main panel, with the associated integrated pulse profile on the left. 
Bottom plot shows the corresponding intensity-modulation pattern averaged over longitudes, 
clearly showing an evidence for periodic nulls. The dashed line represents the 4$\sigma$ 
detection threshold.} 
\end{figure}

The periodic pseudo-nulls found by HR07 are among the best manifestations of the underlying carousel model
responsible for the drifting subpulse phenomenon in pulsars. We have also checked our data for
their presence, although given its erratic nature (Fig.~1) we were not very optimistic at first.
However, to our surprise we managed to detect them in some selected, brighter pulse sequences. 
A good example from our results can be seen in Fig.~4, which shows the pattern of the modulation 
of the pulsar signal, folded with a period of $P_4=28.44 P_1$ found by HR07. Each folding 
was performed over 512 pulses,
i.e. about 18 putative periods $P_4$. In four of such foldings (out of 16), we found a clear
evidence for periodic pseudo-nulls in the form of apparent ``null zones" in the folded modulation
pattern. One of our best examples, as shown in Fig.~4, can be compared with Fig.~7 in HR07. One could
argue, that given the erratic nature  of our data, such nulls can appear in our folded profiles by
pure coincidence, as we do not detect the majority of the incoming pulses. But given the fact that
we detect about 20\% of pulses, the chance of appearance of  such coincidental periodic genuine
nulls (as opposed to periodic pseudo-nulls) is only $0.8^{18}=1.8$\% (to minimize such chance
occurrence we were forced on folding of a very long pulse sequences). Not to mention the fact that the
folding procedure itself helps to detect pulses that individually remain just below our detection
threshold. Indeed, during folding of 18 sequences, we improve our sensitivity by a factor of more
than four ($\sqrt{18}$). Therefore, we believe that the detection of periodic pseudo-nulls in our
8.35 GHz data confirms directly the findings of HR07 at the much lower frequency of 327 MHz. At the
same time, it ultimately proves the carousel model for PSR~B1133+16, that is 23 subbeams completing
one full circulation in exactly $28.44 P_1$. It is worth emphasizing that we detected our
pseudo-nulls at exactly this period, and the LFF in our best
data sequences is also close to this value (Tables 1 and 2).

Summarizing, PSR~B1133+16 seems to be a case of a pulsar that shows an organized behaviour 
of its individual pulses only for some limited-duration periods of time, and otherwise appears 
to be highly chaotic. We still do not know what triggers the switch from chaos to order, and we 
do not undrestand what is happening with the pulsar during the chaotic periods - this definitely requires
further study. However, we seem to have enough evidence to claim that, when the order appears, 
the pulsar radiation can be explained by means of the rotating carousel model. Moreover, our
results presented in this paper show that the most fundamental feature of the carousel
model, that is the carousel circulation time ($P_4$) prevails despite various difficulties
nature (in form of details and nuances of pulsar emission) puts in our way. Whether this is
some aliasing-smeared drifting subpulse phenomenon observed at lower frequencies, or low flux density at
high frequencies preventing the majority of individual pulses from reaching the detection
threshold, the low frequency feature is often visible. Additionaly, one of the most representative 
features of the carousel model is the phenomenon of periodic pseudo-nulls, and it is very important to note 
that we detected them at exactly the value of $28.44 P_1$ reported by HR07 at lower frequencies.

\section*{Acknowledgments}
This paper is based on observations with the 100-m telescope of the MPIfR (Max-Planck-Institut
f\"{u}r Radioastronomie) at Effelsberg. Sneha Honnappa acknowledges the support of the Polish Grant
N~N203~406239. SH, JG, JK, WL and OM  acknowledge the support of the Polish Grant N N203 391934. We
thank K. Krzeszowski and M. Sendyk for technical assistance, and M. Serylak for fruitful discussion.

\end{document}